\documentclass[prd,twocolumn,showpacs,preprintnumbers,aps,nofootinbib]{revtex4-1}

\usepackage{graphicx}
\usepackage{dcolumn}
\usepackage{bm}
\usepackage{amsmath,amssymb,amsfonts}
\usepackage{latexsym}
\usepackage{color}

\def\nn    {\nonumber}

\def\dcp{\Delta \mathcal{A}_{\text{CP}}}
\def\ebsg{B\to X_s \gamma}
\def\bsg{\mathcal{B}(B\to X_s \gamma)}
\def\rbb{\rho_{bb}}
\def\rtt{\rho_{tt}}
\def\irbb{\mbox{Im}(\rho_{bb})}

\def\fbi{fb$^{-1}$\;} 
\def\cba{c_{\beta-\alpha}}
\def\sba{s_{\beta-\alpha}}
\def\bZH{$bg\to b A \to b Z H$\;}
\def\bbZH{$gg \to b \bar b A \to b \bar b Z H$\;}

\begin{document}

\title{\boldmath  {Probing additional bottom Yukawa coupling via $bg\to b A \to b Z H$ signature}}
\author{Tanmoy Modak}
\affiliation{Department of Physics, National Taiwan University, Taipei 10617, Taiwan}
\bigskip

% \date{\today}

\begin{abstract}
The recent discovery of the bottom quark Yukawa coupling ($hbb$) of the 125 GeV scalar motivates one to search for extra bottom
Yukawa coupling that may exist in nature. The two Higgs doublet model without a discrete $Z_2$ symmetry 
allows the possibility of additional bottom Yukawa coupling $\rho_{bb}$. We show that $\rho_{bb}$ can be searched directly at the
LHC via $bg\to b A \to b Z H$ and  $gg \to b \bar b A \to b \bar b Z H$ processes, where $A$ and $H$ are the CP-odd and CP-even scalars
respectively. We find that the $bg\to b A \to b Z H$ process could be discovered with $\sim300$ fb$^{-1}$ integrated
luminosity if $ m_A \sim 300$ GeV, while the latter process may emerge in the high luminosity LHC (HL-LHC) run.
A discovery might touch upon the parameter space required for the electroweak baryogenesis.
\end{abstract}

\maketitle

%-----------------------------------------------------------------------------
%	Introduction
%-----------------------------------------------------------------------------
%%%%%%%%%%%%%%%%%%%%%%%%%%%%%%%%%%%%%%%%%%%%%%%%%%%%%%%%%%%%%%%%%%%%%%%
\section{Introduction}
%%%%%%%%%%%%%%%%%%%%%%%%%%%%%%%%%%%%%%%%%%%%%%%%%%%%%%%%%%%%%%%%%%%
The discovery of the 125 GeV scalar boson $h$~\cite{h125_discovery} and its properties 
corroborate that the Standard Model (SM) is the correct effective theory at around weak scale.
Even though no clear evidence of new physics (NP) has been found, 
the Run-2 era of LHC witnessed one of the most intriguing \textit{discovery}, 
that is the bottom quark Yukawa coupling $hbb$~\cite{Aaboud:2018zhk,Sirunyan:2018kst}.  
The observation was announced simultaneously by the ATLAS and CMS collaborations. 
Both the experiments performed searches mainly in the process where $h$ is
produced in association with a $Z$ or $W$ boson, followed by the $h \to b\bar b$ decay. 
When combined with the results from the other searches of Run-1 and Run-2,
the observed signal strengths relative to the SM expectation were reported to be 
$1.01\pm 0.12(\text{stat.})^{+0.16}_{-0.15}(\text{syst.})$ at ATLAS~\cite{Aaboud:2018zhk}, while
$1.04\pm 0.14(\text{stat.})\pm 0.14(\text{syst.})$ at CMS~\cite{Sirunyan:2018kst}.
Although they are consistent with the SM prediction, both the measurements are quite accommodating for NP contribution.
In the backdrop of these recent observations, it is timely to ask whether there exists any additional bottom Yukawa coupling in nature.
In this article we explore the possibility of direct detection and identification of such extra bottom Yukawa coupling at the LHC.

The context is the two Higgs doublet model (2HDM).
In the absence of discrete $Z_2$ symmetry, which was invoked to ensure Natural Flavor Conservation (NFC)~\cite{Glashow:1976nt} 
to forbid flavor changing neutral Higgs couplings, 
both the doublets couple to up- and down-type quarks.
After the diagonalization of the fermion mass matrices 
two independent Yukawa matrices $\lambda_{ij}^F =({\sqrt{2}m_i^F}/{v})\, \delta_{ij}$  
(with $v \simeq 246$ GeV) and $\rho_{ij}^F$ emerge, where $F$ denotes up- and down-type quarks and, leptons. 
The Yukawa matrices $\lambda_{ij}^F$ are real and diagonal, where as, $\rho_{ij}^F$ are in general non-diagonal and complex. 
Our focus of interest is the the extra bottom Yukawa coupling $\rho_{bb}$. 
In this paper, we analyze the prospect its direct detection at the LHC
via \bZH and \bbZH processes (charge conjugate processes are implied) with $b$-tagging.

We investigate the discovery potential of $\rbb$ via $pp\to b A + X \to b ZH + X$ ($X$ is inclusive activity) with
$Z\to \ell^+ \ell^-$ ($\ell = e, \mu$) and $H\to b \bar b$ (denoted as $bZH$ process) at the $14$ TeV LHC.
In finding the discovery potential we assumed the extra top Yukawa coupling $\rtt$ to be relatively 
small to avoid the direct search constraints from $gg\to A/H$.
A sizable $\rbb$ would also induce  \bbZH which provides additional probe for $\rbb$. 
We study this process via $pp\to b \bar b A + X \to b \bar b ZH + X$ followed by
$Z\to \ell^+ \ell^-$ and $H\to b \bar b$ (denoted as $bbZH$ process). 
Recently, $\rho_{bb}$ received additional significance as it can drive 
electroweak baryogenesis (EWBG) rather efficiently~\cite{Modak:2018csw}. 
It was shown that $\mathcal{O}(0.1)$  imaginary $\rho_{bb}$ ($\irbb$) 
can successfully generate the observed Baryon Asymmetry of the Universe~\cite{Modak:2018csw}.
Although, the information of the phase could not be captured in the $bZH$ and $bbZH$ processes, 
however, a discovery might indicate $\rbb$ driven EWBG.

The paper is organized as follows. We outlined the formalism in the Sec.~\ref{form}, followed by discussion on the relevant constraints
and available parameter space in the Sec.~\ref{paramspace}. The Sec.~\ref{disc} is dedicated to the collider signatures of 
the $bZH$ and $bbZH$ processes respectively. We summarized our results with some discussions in the Sec.~\ref{summ}.

%%%%%%%%%%%%%%%%%%%%%%%%%%%%%%%%%%%%%%%%%%%%%%%%%%%%%%%%%%%%%%%%%%%%%%%%%%%%%%%
\section{Formalism}\label{form}
%%%%%%%%%%%%%%%%%%%%%%%%%%%%%%%%%%%%%%%%%%%%%%%%%%%%%%%%%%%%%%%%%%%%%%%%%%%%%%%
The most general $CP$-conserving two Higgs doublet potential is
given in the general basis as~\cite{Davidson:2005cw, Hou:2017hiw}
\begin{align}
 & V(\Phi,\Phi') = m_{11}^2|\Phi|^2 + m_{22}^2|\Phi'|^2 - (m_{12}^2\Phi^\dagger\Phi' + h.c.)
 \nn\\
 & \quad + \frac{\lambda_1}{2}|\Phi|^4 + \frac{\lambda_2}{2}|\Phi'|^4 + \lambda_3|\Phi|^2|\Phi'|^2
              + \lambda_4 |\Phi^\dagger\Phi'|^2\nn\\
 & + \bigg[\frac{\lambda_5}{2}(\Phi^\dagger\Phi')^2
     + \left(\lambda_6 |\Phi|^2 + \lambda_7|\Phi'|^2\right) \Phi^\dagger\Phi' + h.c.\bigg],
\label{pot}
\end{align}
where the parameters $m_{11}^2$, $m_{12}^2$, $m_{22}^2$ and $\lambda_i$s are all real.
The vacuum expectation values of the doublets $\Phi$ and $\Phi'$ are given as $\left\langle \Phi\right\rangle =(0,v_1/\sqrt{2})^T$ and 
$\left\langle \Phi'\right\rangle =(0,v_2/\sqrt{2})^T$ respectively, such that $v^2 = v_1^2 + v_2^2= (246~\mbox{GeV})^2$ and $\tan \beta =v_2/v_1$. 
With no $Z_2 $ symmetry in place to distinguish between $\Phi$ and $\Phi'$, $\tan\beta$ becomes unphysical. We move to the Higgs basis 
through basis rotation such that $\left\langle \Phi\right\rangle =(0,v/\sqrt{2})^T$ and $\left\langle \Phi'\right\rangle =(0,0)^T$,
where the parameters in the Higgs basis can be identified by the replacements $m_{ij}^2 \to \mu_{ij}^2$ and $\lambda_i \to \eta_i$
~\footnote{The relations between the parameters in the two bases can be found out in Ref.~\cite{Davidson:2005cw}.}.
The scalar potential minimization conditions lead to $\mu_{11}^2=-\frac{1}{2}\eta_1 v^2$ and 
$\mu_{12}^2=\frac{1}{2}\eta_6 v^2$, with $\mu_{22}^2 > 0$.
The mixing angle $\beta-\alpha$ between the $CP$ even bosons $h$ and $H$
satisfies the {relations}~\cite{Hou:2017hiw}
\begin{align}
 \cba^2 = \frac{\eta_1 v^2 - m_h^2}{m_H^2-m_h^2},~\quad  \sin{2({\beta-\alpha})} = -\frac{2\eta_6 v^2}{m_H^2-m_h^2},
\end{align}
with shorthand $\cos(\beta-\alpha)=\cba$ and $\sin(\beta-\alpha)=\sba$.
In the alignment limit $\cba=0$, which leads to $\eta_1 v^2 = m_h^2$.
The  masses of the charged scalar $m_{H^\pm}$ and 
the neutral scalars $A$, $h$ and $H$ can be expressed as:
\begin{align}
 &m_{H^\pm}^2 = \frac{1}{2}\eta_3 v^2+ \mu_{22}^2,\\
 &m_{A}^2 = \frac{1}{2}(\eta_3 + \eta_4 - \eta_5) v^2+ \mu_{22}^2,\\
 &m_{h,H}^2 = \frac{1}{2}\bigg[m_A^2 + (\eta_1 + \eta_5) v^2\nn\\
 &\quad\quad \quad\quad \mp \sqrt{\left(m_A^2+ (\eta_5 - \eta_1) v^2\right)^2 + 4 \eta_6^2 v^4}\bigg].
\end{align}

The $CP$-even scalars $h$, $H$, $CP$-odd scalar $A$ and, the charged scalar $H^\pm$ couple to the fermions 
by~\cite{Davidson:2005cw}
\begin{align}
\mathcal{L}_{\small{Y}}=&-\frac{1}{\sqrt{2}} \sum_{F = U, D, L}
 \bar F_{iL} \bigg[\big(\lambda^F_{ij} \sba + \rho^F_{ij} \cba\big) h \nn\\
 &+\big(\lambda^F_{ij} \cba - \rho^F_{ij} \sba\big)H -i ~{\rm sgn}(Q_F) \rho^F_{ij} A\bigg] F_{jR}\nn\\
 &-\bar{U}_i\left[(V\rho^D)_{ij} P_R-(\rho^{U\dagger}V)_{ij} P_L\right]D_j H^+ \nn\\
 &- \bar{\nu}_i\rho^L_{ij} P_R L_j H^+ +{\rm h.c.},\label{eff}
\end{align}
where $i,j =1,2,3$ are the generation  indices that are summed over, $P_{L,R}=\frac{1\mp\gamma_5}{2}$, $V$ is the CKM matrix, 
$\lambda^F_{ij}$ are real and diagonal and $\rho^F_{ij}$ are complex non-diagonal $3\times 3$ matrices.

Our process of interest is \bZH, where the production process is initiated by $\rbb$ and 
the decay $A \to Z H$ is conformed by
\begin{align}
 &\frac{g_2 }{2 c_W}Z_\mu \bigg(\cba(h\partial^\mu A 
 - \partial^\mu h \cdot A)\nn\\
 &-\sba(H\partial^\mu A
 - \partial^\mu H \cdot A)\bigg),\label{zhlagra}
\end{align}
with $g_2$ is the $SU(2)_L$ gauge coupling and $c_W$ is Weinberg's angle.
% For $m_A + m_Z < m_H$, $\rbb$ would invoke $bg\to b H \to b Z A$, however, unlike \bZH, the discovery potential is 
% suppressed due to $H \to hh$ decay, if $m_H > 2 m_h$ and $\cba$ is non-zero. 
A search can be performed in $bg\to b A\to b \bar b  b$ mode, but, the process suffers from
the overwhelming QCD multijets backgrounds. For $m_{H^\pm}+  m_W < m_A$ ($m_W$ is the $W$ boson mass)
$bg\to b A \to b H^+ W^-$ is possible, but if searched in $H^+ \to t \bar b$ (induced by $\rbb$) 
with $t \to b \ell^+ \nu_\ell$ one loses the mass reconstruction
capability of $A$ and, hence, controlling of the $t\bar t$ background.
In general, backgrounds are even higher if searched in the hadronically decaying $t$ mode. 
Notice that, the $\rtt$ coupling, which also can induce $H^+ \to t \bar b$ decay,
obfuscates the role of $\rbb$. We remark that the \bZH process, which can only be induced via $\rbb$, offers a unique probe 
for the $\rbb$ coupling~\footnote{In principle $qg \to b A \to b Z H$ can replicate the same final 
state as in \bZH process in the  $pp$ collision
if $\rho_{bq}$ or $\rho_{qb}$ ($q= d$ and $s$) are sizable,
however, they receive severe constraints from the $B_q$ mixing~\cite{Chen:2018hqy}.}.

The process $bg\to b A \to b Zh$ is indeed possible, however
suppressed by the mixing angle $\cba$. 
It should be clear from Eq.~\eqref{zhlagra} that the decay $A\to ZH$ is proportional to $\sba$. 
As a result, a discovery of \bZH process is possible  even in the approximate alignment (i.e. for small $\cba$),
which is observed at the LHC~\cite{approxalign}.
Further, $\rbb$ can initiate $b\bar b \to  A \to Z H$ and loop induced $gg\to A \to Z H$~\cite{gg2ZH}, 
however, the coupling information is lost in the $pp$ collision. Besides,
as $\rho_{tt}$ can also get involved in the loop, the role of $\rho_{bb}$ is obscured in $gg\to A \to Z H$.
One can also have $gg\to A \to Z h $~(see e.g. Refs.~\cite{Aaboud:2017cxo,Sirunyan:2019xls} and references therein)
and $gg\to b \bar b A \to b\bar b Z h$ (see e.g. Refs.~\cite{Aaboud:2017cxo,Sirunyan:2019xls,Ferreira:2017bnx,Coyle:2018ydo}), 
however, again both processes are suppressed by the mixing angle $\cba$ .

%%%%%%%%%%%%%%%%%%%%%%%%%%%%%%%%%%%%%%%%%%%%%%%%%%%%
\section{Allowed parameter space}\label{paramspace}
%%%%%%%%%%%%%%%%%%%%%%%%%%%%%%%%%%%%%%%%%%%%%%%5
Having already set up the formalism we now focus on the relevant constraints and the available parameter space for our study. 
We first scrutinize the constraints on $\rho_{bb}$. For simplicity, we set all $\rho_{ij}=0$ except for $\rbb$ and $\rtt$ in this section.
We assume small $\rtt$ in order to avoid direct search limits from $gg\to A/H$. 
In particular we choose $\rtt = 0.1$ for illustration. The most stringent constraints arise from the Higgs signal strength measurements,
the branching ratio of $B\to X_s \gamma$ ($\bsg$), the asymmetry of the CP asymmetry 
between the charged and neutral $B\to X_s \gamma$ decays ($\dcp$), electron electric dipole moment (EDM) measurement and
the upper limit on the $h$ decay width.

\begin{figure*}[htb!]
\center
\includegraphics[width=.4 \textwidth]{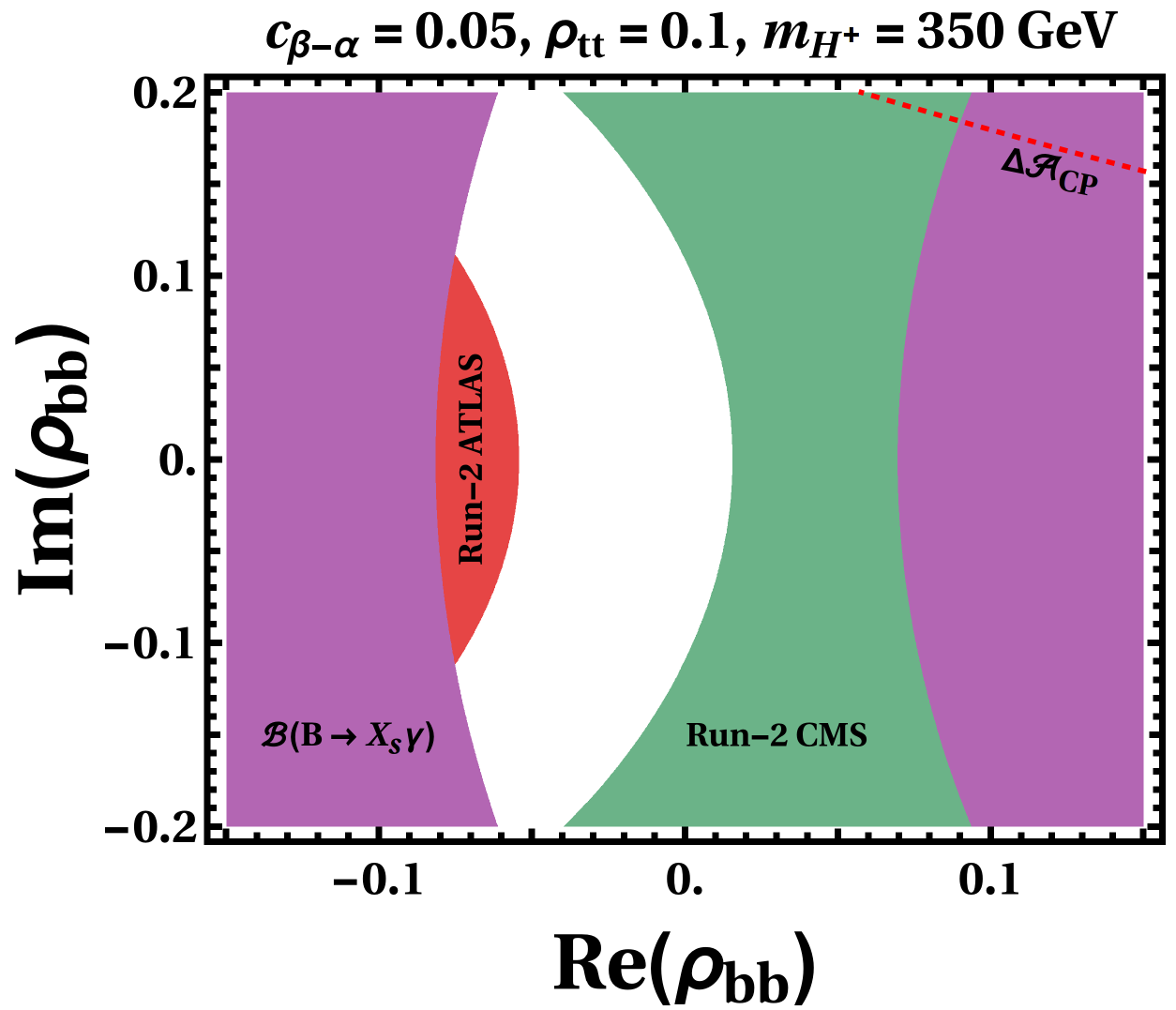}
\includegraphics[width=.4 \textwidth]{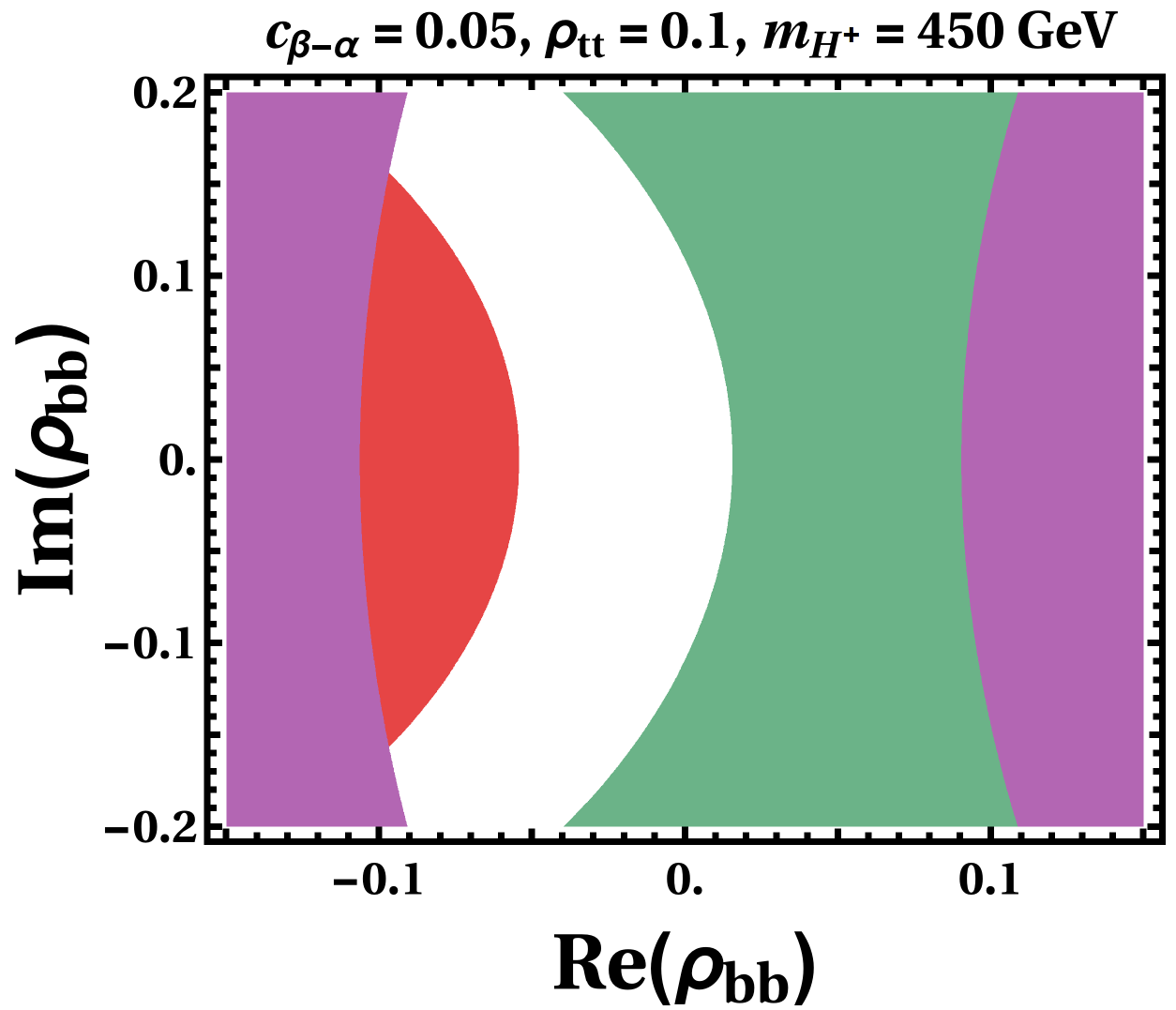}
\caption{The constraints on $\rbb$ from Higgs signal strength measurements 
of ATLAS (red shaded region), Higgs signal strength measurements of 
CMS (green shaded region), $\bsg$ (purple shaded region)
and $\dcp$ (red dotted line). The figures are generated with 
$\cba = 0.05$ and $\rtt = 0.1$, for $m_{H^\pm} = 350$ GeV (left panel)
and 450 GeV (right panel). See text for further details.} 
\label{const}
\end{figure*}

The couplings $\rho_{ij}$ modify the $h$ boson couplings to the fermions 
for moderate values of $\cba$, as can be seen from  Eq.\eqref{eff}. Therefore, $\rho_{bb}$ receives meaningful constraint from the
Higgs boson coupling measurements, unless $\cba$ is vanishingly small. 
We utilized Run-2 ATLAS~\cite{ATLAS:2019slw} and CMS~\cite{Sirunyan:2018koj} measurements 
which are based on 80 \fbi and 35.9 \fbi data respectively.
The results summarize the values of different signal strengths $\mu_i^f$ and corresponding errors
to a particular decay mode $i\to h \to f$. 
Following the Refs.~\cite{ATLAS:2019slw,Sirunyan:2018koj}, we define a signal strength $\mu_i^f$ as:
\begin{align}
 \mu_i^f = \frac{\sigma_i\mathcal{B}^f}{(\sigma_i)_{\text{SM}}(\mathcal{B}^f)_{\text{SM}}} = \mu_i \mu^f,
\end{align}
where $\sigma_i$ is denoted as the production cross section of $i\to h$ and $\mathcal{B}^f$ is the branching ratio
for $h\to f$. The production modes considered are $i=ggF~(\mbox{gluon-fusion})$,$~VBF~(\mbox{vector-boson-fusion})$, $Zh$, $Wh$, $tth$, 
and the branching ratios are $f= \gamma\gamma,~ZZ,~WW,~\tau\tau,~bb,~\mu\mu$. 
For simplicity we utilized the LO $\mu_i^f$ in our analysis and followed Refs.~\cite{Djouadi:2005gi,Branco:2011iw,Fontes:2014xva,Hou:2018uvr} for their 
explicit expressions. In particular, we focused on two different production modes, the $ggF$ and the $VBF$
in our analysis. We find that for the $ggF$ category, the most relevant signal strengths for our analysis are
$\mu_{ggF}^{ZZ}$, $\mu_{ggF}^{WW}$, $\mu_{ggF}^{\gamma\gamma}$ and $\mu_{ggF}^{\tau\tau}$, while in the $VBF$ category 
$\mu_{VBF}^{WW}$, $\mu_{VBF}^{\gamma\gamma}$ and $\mu_{VBF}^{\tau\tau}$.
% The values and corresponding errors of these signal strengths
% can be found out in Table.~3 of Ref.~\cite{Sirunyan:2018koj}. 
In addition, we further considered the recent observation of the $h\to b\bar b$ in the 
$Vh$ production by ATLAS~\cite{Aaboud:2018zhk} and CMS~\cite{Sirunyan:2018kst}. 
We referred them together as ``Higgs signal strength measurements''. 
The parameter space excluded by the Higgs signal strength measurements are 
shown by the red (ATLAS) and green (CMS) shaded regions in Fig.~\ref{const}
for $m_{H^\pm} = 350$ GeV (left) and 450 GeV (right). 
In generating Fig.~\ref{const}, we allowed $2\sigma$ errors on each signal strength measurements and, assumed $\cba=0.5$.

The branching ratio measurement of $B \to X_s \gamma$ provide another stringent constraint on $\rbb$. The coupling $\rbb$ enters in the $\bsg$  
via charged Higgs and top quark loop. At the matching scale $\mu = m_W$, the modified leading order
(LO) Wilson coefficients $C^{(0)}_{7,8}$ are defined as
\begin{align}
C^{(0)}_{7,8}(m_W)= F^{(1)}_{7,8}(x_t)+\delta C_{7,8}^{(0)}(\mu_W),\label{c78}
\end{align}
with, $\overline{m}_t(m_W)$ is the $\overline{\mbox{MS}}$ running mass of top quark at $m_W$, $x_t=(\overline{m}_t(m_W)/m_W)^2$, 
while the expression for $F^{(1)}_{7,8}(x)$ can be found out in the Refs.~\cite{Ciuchini:1997xe,Chetyrkin:1996vx}. 
The second term in Eq.\eqref{c78}, which originates from 
the charged Higgs contribution, expressed at LO as~\cite{Altunkaynak:2015twa}
\begin{align}
 \delta C_{7,8}^{(0)}(m_W)\simeq &\frac{|\rtt|^2}{3\lambda_t^2}F^{(1)}_{7,8}(y_{H^+}) 
 -\frac{\rtt\rho_{bb}}{\lambda_t\lambda_b}F^{(2)}_{7,8}(y_{H^+}),
\end{align}
where $y_{H^+}=(\overline{m}_t(m_W)/m_{H^+})^2$. Here we have followed Ref.~\cite{Ciuchini:1997xe}, for the expression of $F^{(2)}_{7,8}(y_{H^+})$.
The current world average of $\bsg_{\text{exp}}$, which is extrapolated to the photon energy cut $E_0=1.6$ GeV 
is found to be  $(3.32\pm0.15)\times 10^{-4}$~\cite{Amhis:2016xyh}. The
SM prediction of $\bsg$ at next-to-next-to LO (NNLO) for the same photon 
energy cut is $(3.36\pm0.23)\times 10^{-4}$~\cite{Czakon:2015exa}.
In order to find the constraint on $\rho_{bb}$, we adopted the prescription of Ref.~\cite{Crivellin:2013wna} and defined
\begin{align}
 R_{\text{exp}}=\frac{\bsg_{\text{exp}}}{\bsg_{\text{SM}}}.
\end{align}
Based on our LO calculation we further defined
\begin{align}
 R_{\text{theory}}=\frac{\bsg_{\text{G2HDM}}}{\bsg_{\text{SM}}},
\end{align}
and took $m_W$ and $\overline{m}_b(m_b)$ as the matching scale and low-energy scales respectively.
We finally demanded $R_{\text{theory}}$ should not exceed $2\sigma$ error of $R_{\text{exp}}$. The excluded regions
are displayed by the purple shaded regions in Fig.~\ref{const}.

The direct CP asymmetry $\mathcal{A}_{\text{CP}}$~\cite{Kagan:1998bh} of $B\to X_s \gamma$ is sensitive to $\irbb$.
However, it has been proposed~\cite{Benzke:2010tq} that $\dcp$,
defined as the asymmetry of the CP asymmetry between the charged and neutral $B\to X_s \gamma$ decay 
provides even more powerful probe for the CP violating effects. The $\dcp$ is defined as~\cite{Benzke:2010tq}
\begin{align}
 \dcp = \mathcal{A}_{B^-\to X_s^- \gamma} - \mathcal{A}_{B^0\to X_s^0 \gamma}
 \approx 4 \pi^2 \alpha_s \frac{\tilde{\Lambda}_{78}}{m_b}\mbox{Im}\bigg(\frac{C_8}{C_7}\bigg),\label{acp}
\end{align}
where, $\alpha_s$ is the strong
coupling constant calculated at $\overline{m}_b(m_b)$ and $\tilde{\Lambda}_{78}$ is a hadronic parameter.  
It is expected that hadronic parameter $\tilde{\Lambda}_{78} \sim\Lambda_{\text{QCD}}$
and estimated to be in the range of $ 17~\mbox{MeV}<\tilde{\Lambda}_{78}<190$ MeV~\cite{Benzke:2010tq}.
We take the average value of $\tilde{\Lambda}_{78}=89$ MeV as a reference value for illustration. 
A recent Belle measurement report $\dcp=(+3.69\pm2.65\pm0.76)\%$~\cite{Watanuki:2018xxg}, where the first and second uncertainties are 
statistical and systematic respectively. Utilizing Eq.~\eqref{acp} and allowing $2\sigma$ error on 
the Belle measurement of $\dcp$ we find the  red dotted lines (the regions above are excluded) in Fig.~\ref{const} 
for $m_{H^\pm} = 350$ GeV and 450 GeV.
As a first approximation we have utilized the LO Wilson coefficients as in Eq.~\eqref{c78} in our analysis.
We stress that the constraint heavily depends on the value of $\tilde{\Lambda}_{78}$ and becomes 
stronger for the larger values of $\tilde{\Lambda}_{78}$.

\begin{figure*}[htb]
\center
\includegraphics[width=.4 \textwidth]{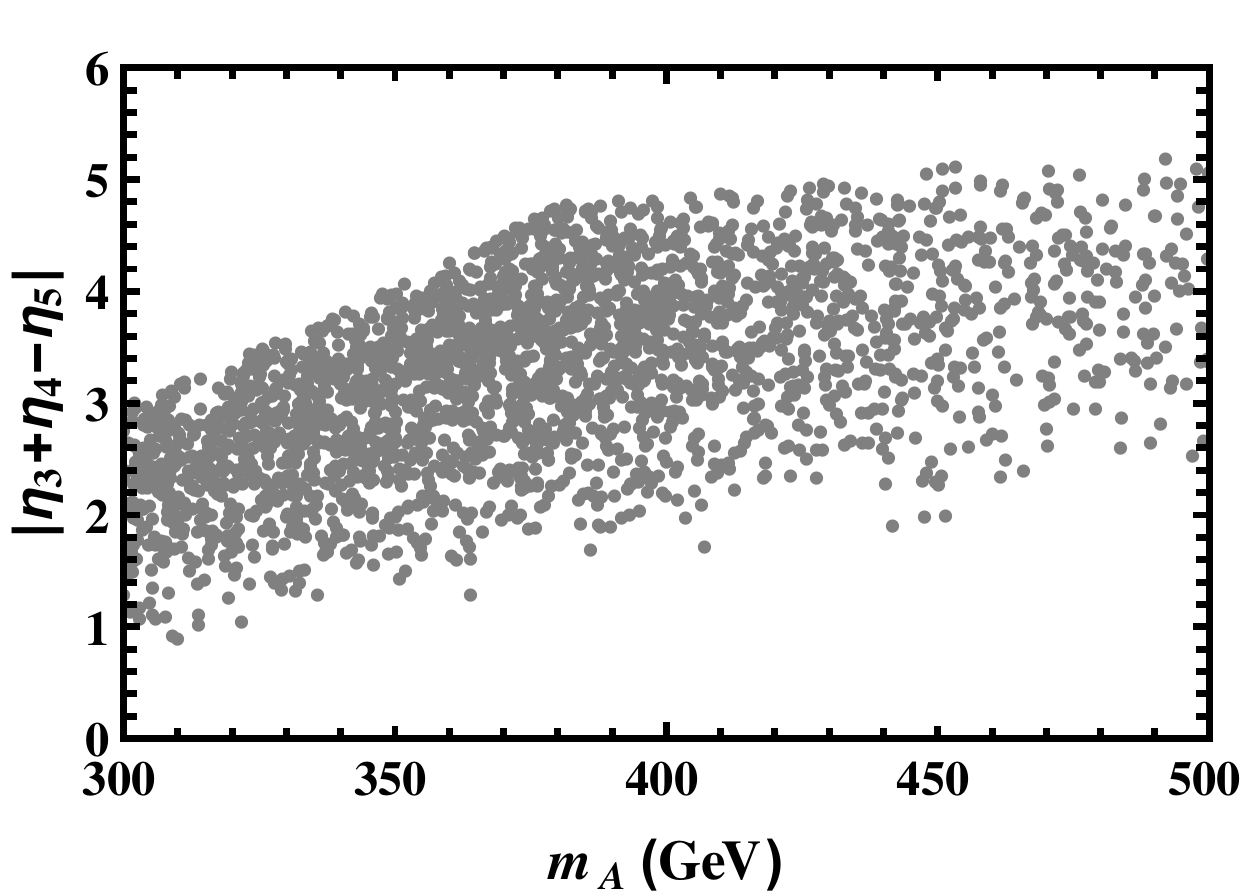}
\includegraphics[width=.4 \textwidth]{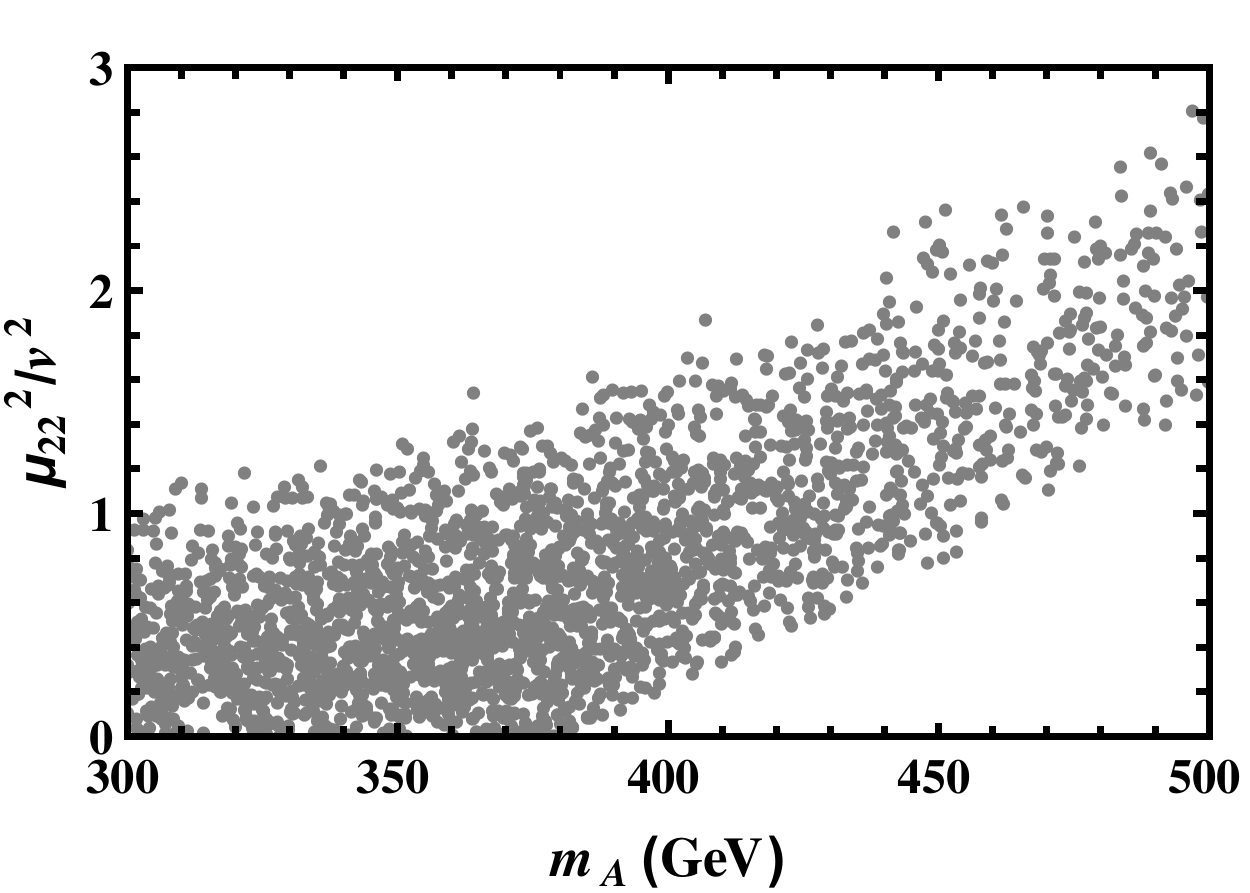}
\caption{The scanned points plotted in the $|\eta_3+\eta_4 -\eta_5|$ vs $m_A$ (left) and $\mu_{22}^2/v^2$ vs $m_A$  (right)  plane.} 
\label{scanplot}
\end{figure*}

The most stringent constraint on $\irbb$ comes from the electron EDM ($d_e$) measurements. 
The  two-loop Barr-Zee diagrams~\cite{Barr:1990vd}, which is studied widely in the context of 2HDM~\cite{EDM_2HDM}, 
are the leading contributions to $d_e$.
A recent result from ACME Collaboration finds $|d_e|<1.1\times 10^{-29}~e~\text{cm}$~\cite{Andreev:2018ayy}, which 
excludes even the nominal value (i.e. $|\irbb|\lesssim 0.058$) required for $\rbb$ driven EWBG~\cite{Modak:2018csw}.
The constraint could be relaxed by either turning on $\rho_{ee}$, or even could vanish in the alignment limit.
In the former scenario non-zero $\rho_{ee}$ and $\rbb$ induce other Barr-Zee diagrams with opposite sign, where as
in the  the latter case all the contributions to the EDM are simply decoupled.
In particular, Ref.~\cite{Modak:2018csw} finds for $\text{Re}(\rho_{ee})=0$, $\mathcal{O}(0.1)$ $\irbb$ is still allowed 
if $0.06\lesssim \text{Im}(\rho_{ee})/(\lambda_e\lambda_b)\lesssim 0.3$.

The current upper limit of the $h$ boson decay width, which is extracted to be $< 0.013$ GeV 
(95\% CL)~\cite{Tanabashi:2018oca}, can provide some limit on $|\rbb|$ if $\cba \neq 0$. 
Besides, the presence of the additional scalars modify the $Zbb$ vertex~\cite{Haber:1999zh} at one-loop and in principle can constrain
the $\rbb$ coupling. However, we found these limits to be weaker and 
lie beyond the plotted ranges in Fig.~\ref{const}.

Let us understand Fig.~\ref{const}.
In generating Fig.~\ref{const} we set $\cba = 0.05$ and $\rtt=0.1$ for illustration.
The constraint from the Higgs signal strength measurements depend primarily on the value of $\cba$ and 
vanish in the alignment limit. Same is true for the constraint from the electron EDM, which also disappears in the alignment
limit. On the other hand, bounds from  $\bsg$ and $\dcp$ alleviate for small $\rtt$, and/or for heavier $H^\pm$. 
It is clear from Fig.~\ref{const} that $|\rbb|\sim 0.1$ is still allowed, however, $\cba$ and $\rtt$ should not be very large. In the following
we would assume the alignment limit and set $\rtt=0$ for the sake of simplicity, 
however their impacts will be discussed in the latter part of the paper.
In passing we remark that there exist several direct searches at the LHC which can also constrain
$\rbb$, even for $\cba =0$ and $\rtt =0$. We defer a detailed discussion of them for the next section.

\begin{table*}[hbt!]
\centering
\begin{tabular}{c |c| c| c| c | c | c| c | c |c| c| c}
\hline
BP  & $\eta_1$ &  $\eta_2$   &  $\eta_3$   & $\eta_4$  & $\eta_5$  &\hspace{0.1cm} $\eta_6$ \hspace{0.1cm} & $\eta_7$  & $m_{H^\pm}$  & $m_A$ & $m_H$ 
  & $\frac{\mu_{22}^2}{v^2}$\\
 &&&&&&& & (GeV) & (GeV) &(GeV)  &\\ 
 &&&&&&&&&&&\\
\hline
\hline
   &&&&&&&&&&&\\
I                     & 0.258 & 1.151 & 2.78  & -1.557 & -0.831 & 0 & -0.236 & 335 & 301 & 200  & 0.46 \\

II                    & 0.258 & 2.155 & 2.496 & -0.625 & -1.885 & 0 & 0.375  & 349 & 400 & 214 & 0.76 \\

III                    & 0.258 & 2.63  & 2.142 & -0.333 & -2.436 & 0 & -0.134 & 426 & 495 & 312 & 1.92 \\

 &&&&&&&&&&&\\
\hline
\hline
\end{tabular}
\caption{Parameter values for the three benchmark points. See text for details.}
\label{bench}
\end{table*}

For the dynamical parameters in Eq.~\eqref{pot}, 
one needs to satisfy the perturbativity, positivity and tree-level unitarity conditions, for which we utilized 2HDMC~\cite{Eriksson:2009ws}.
The quartic couplings $\eta_1$, $\eta_{3{\rm -}6}$ can be expressed in 
terms of $m_h$, $m_A$, $m_H$, $m_{H^\pm}$, $\mu_{22}$ and mixing angle ${\beta-\alpha}$,
all normalized to $v$~\cite{Hou:2017hiw}:
\begin{align}
 & \eta_1 = \frac{m_h^2 s_{\beta-\alpha}^2 + m_H^2 c_{\beta-\alpha}^2}{v^2},\\
 & \eta_3 =  \frac{2(m_{H^\pm}^2 - \mu_{22}^2)}{v^2},\\
%  & \eta_4 =  \frac{2(m_A^2 - \mu_{22}^2)}{v^2}-\eta_3+\eta_5,\\
 & {\eta_4 = \frac{m_h^2 c_{\beta-\alpha}^2 + m_H^2 s_{\beta-\alpha}^2 -2 m_{H^\pm}^2+m_A^2}{v^2}},\\
& \eta_5 =  \frac{m_H^2 s_{\beta-\alpha}^2 + m_h^2 c_{\beta-\alpha}^2 - m_A^2}{v^2},\\
 & \eta_6 =  \frac{(m_h^2 - m_H^2)s_{\beta-\alpha}c_{\beta-\alpha}}{v^2}.
\end{align}
The mixing angle ${\beta-\alpha}$ and the quartic couplings $\eta_2$ and  $\eta_7$ are not related to masses. 
Hence, we take $v$, $m_h$, $m_A$, $m_H$, $m_{H^\pm}$, $\cba$, $\eta_2$, $\eta_7$ and $\mu_{22}$
as the phenomenological parameters.
However, in order to save computation time, we randomly generated these parameters 
in the following ranges:
$\mu_{22} \in [0, 700]$  GeV,
$m_A \in [300, 500]$ GeV,
$m_{H^\pm} \in [300, 800]$ GeV,
$\eta_2 \in [0, 3]$, $ \eta_7 \in [-3, 3]$, $m_H \in [200, m_A - m_Z]$ GeV, 
and $\cba = 0$ while satisfying $m_h = $  125 GeV.
Further, we demanded $m_A < m_{H^\pm}+m_W$ to forbid the $A\to H^\pm W^\mp$ decay for simplicity. 
In general, heavier $m_A$ is possible, however the discovery potential 
would alleviate due to the rapid fall in the parton luminosity.
These randomly generated parameters are then passed to 2HDMC, which uses the input parameters~\cite{Eriksson:2009ws} 
$m_{H^\pm}$ and $\Lambda_{1-7}$ in the Higgs basis, and with $v$ as implicit parameter while scanning. 
We identify $\Lambda_{1-7}$ with $\eta_{1-7}$ and take $-\pi/2\leq \beta-\alpha \leq \pi/2$ to match the convention of 2HDMC.
We further conservatively require all $|\eta_i| \leq 3$, however,  
$\eta_2 >0$ is demanded by the positivity of the potential in Eq.~\eqref{pot}, in addition to other
involved conditions in 2HDMC.

We further imposed the stringent oblique $T$ parameter~\cite{Peskin:1991sw} constraint,
which restricts the scalar masses $m_H$, $m_A$ and $m_{H^\pm}$~\cite{Froggatt:1991qw,Haber:2015pua}, 
and hence $\eta_i$s. Utilizing the expression given in Ref.~\cite{Haber:2015pua} the points that
passed unitarity, perturbativity and positivity conditions from 2HDMC,
were further required to satisfy the $T$ parameter constraint within the $2\sigma$ error~\cite{Baak:2013ppa}.
These points are denoted as ``scanned points''.
Finally the scanned points are plotted as gray dots in Fig.~\ref{scanplot} in 
the $|\eta_3+\eta_4 -\eta_5|$ vs $m_A$ and $\mu_{22}^2/v^2$  vs $m_A$ plane.
The Fig.~\ref{scanplot} implies that finite parameter space exist for $300~\mbox{GeV}\lesssim m_A \lesssim 500$ GeV
~\footnote{See also Ref.~\cite{Hou:2019qqi} for more on the parameter counting and scanning strategy.}. 

%%%%%%%%%%%%%%%%%%%%%%%%%%%%%%%%%%%%%%%%%%%%%%%%%%%%%%%%%%%%%%%%%
\section{Collider signatures}\label{disc} 
%%%%%%%%%%%%%%%%%%%%%%%%%%%%%%%%%%%%%%%%%%%%%%%%%%%%%%%%%%%%%%%%%
In this section we analyze the discovery potential of $pp\to b A + X \to b ZH + X$ and $pp\to b \bar b A + X \to b \bar b ZH + X$ 
processes, followed by $H\to b \bar b$ and $Z\to\ell^+\ell^-$ decays. 
In general $Z\to \nu\bar \nu$ and $Z\to \tau^+ \tau^-$ are possible, however, we found these modes to be not as promising. 
In order to illustrate the discovery potential we took three
benchmark points (BPs) from the scanned points in Fig.~\ref{scanplot}, which are summarized in Table~\ref{bench}.
As discussed earlier, the phase information of $\rbb$ is lost in the \bZH and \bbZH processes.
Therefore, the only meaningful quantity in this section is the absolute value of $\rbb$ ($|\rbb|$).
Unless otherwise specified we would only consider $|\rbb|$ from here on.

There exist several direct search limits from ATLAS and CMS that may restrict the 
parameter space of $\rbb$, even for $\cba=0$ and $\rtt=0$.
We find that the searches of heavy Higgs boson, in particular Refs.~\cite{Sirunyan:2018taj,ATLAS:2019jzx,CMS:2018qbg,Aaboud:2018cwk}
are the relevant ones for our study. The most stringent bound arises from the CMS search for 
a heavy Higgs boson production in association with least one additional $b$ 
quark and decaying into $b\bar b$ pair~\cite{Sirunyan:2018taj}. The search is performed with 13 TeV 35.7 \fbi data. 
It sets model-independent 95\% CL upper limits on the $\sigma(pp\to b A/H +X)\cdot\mathcal{B}(A/H\to b \bar b)$ for $m_A/m_H$ ranging from 
300 to 1300 GeV. Utilizing this result we have extracted~\cite{extrac} 95\% CL $\sigma(pp\to b A/H +X)\cdot\mathcal{B}(A/H\to b \bar b)$ upper limit
for BPI, BPII and BPIII. We then calculated the production cross sections ($pp\to b A/H +X $) at the leading order (LO) 
for the three BPs for a reference $|\rbb|$ value using Monte Carlo event generator MadGraph5\_aMC@NLO~\cite{Alwall:2014hca} with 
the default NN23LO1 parton distribution function (PDF) set~\cite{Ball:2013hta}.
Since, CMS does not veto additional activity in the event, we also included contributions from $gg \to b \bar b A/H$ along with
$bg\to b A/H$ in the cross-section estimation. The cross sections are then rescaled by $|\rbb|^2\times\mathcal{B}(A/H\to b \bar b)$
to get the corresponding 95\% CL upper limits on $|\rbb|$. 
The upper limits for the BPI is $|\rbb|\lesssim 0.5$, where as $ |\rbb| \lesssim 0.26$ and $|\rbb|\lesssim 0.24$
for BPII and BPIII respectively. A similar search has been performed by ATLAS~\cite{ATLAS:2019jzx} however the 
limits are somewhat weaker than that of Ref.~\cite{Sirunyan:2018taj}.
The limits extracted from Ref.~\cite{CMS:2018qbg}, which searches for a light scalar decaying into $b\bar b$ pair,
are weaker except for BPI, which we found also to be $|\rbb|\lesssim 0.5$ (at 95\% CL). 
Moreover, ATLAS search for $H^\pm$ in association with a $t$ quark and a $b$ quark with $H^+/H^- \to t \bar b/ \bar t b$ 
decay~\cite{Aaboud:2018cwk} is relevant, but the constraints are milder for all three BPs. 
The effective model is implemented in the FeynRules~2.0~\cite{Alloul:2013bka}.

\begin{table}[hbt!]
\centering
\begin{tabular}{|c| c| c |c |c  c }
\hline
 &&&\\ 
$|\rbb|$ & BP &  \hspace{.3cm}  $ZH$  \hspace{.3cm}&  \hspace{.3cm} $b\bar b$ \hspace{.3cm} \\
&&&\\             
\hline
\hline
                & I          & 0.618 & 0.382  \\
     0.1        & II         & 0.047 & 0.953     \\
                & III        & 0.05  & 0.95   \\            
\hline
\hline
\end{tabular}
\caption{The branching ratios of $A$ for the three benchmark points given in Table~\ref{bench}.}
\label{branch}
\end{table}

We choose $|\rbb| = 0.1$ as a representative value for illustration in this section.
Since our working assumptions are alignment limit with all $\rho_{ij}=0$ except $\rbb$, the total decay width of $A$  is nicely
approximated as the  sum of the partial widths of $A\to b \bar b$ and $A\to Z H$, while $H$ only decays to $b\bar b$.
The corresponding branching ratios of $A$ for the three BPs are given in Table~\ref{bench}, where as $\mathcal{B}(H\to b \bar b)\approx100\%$
for all three BPs. Note, non-zero $\rbb$ induces $A/H\to \gamma\gamma$ and $A/H\to g g$ decays at one loop.
These branching ratios are negligibly small, and hence neglected.

%%%%%%%%%%%%%%%%%%%%%%%%%%%%%%%%
\subsection{The $bZH$ process}
%%%%%%%%%%%%%%%%%%%%%%%%%%%%%%%%
There exist several SM backgrounds for the $bZH$ process.
The dominant backgrounds are $t\bar t+$jets, Drell-Yan+jets (DY+jets), $Wt+$jets, $t\bar tZ$+jets, $t\bar t h$, $tZ$+jets,
with subdominant contributions arise from four-top ($4t$), $t\bar t W$, $tWh$, $tWZ$ and $WZ$+jets. 
Backgrounds from $WW$+jets and $ZZ$+jets are negligibly small and hence not included.
The signal and background event samples are generated at LO, utilizing 
MadGraph5\_aMC@NLO for $pp$ collisions at $\sqrt{s}=14$ TeV
with the PDF set NN23LO1 and then interfaced with PYTHIA~6.4~\cite{Sjostrand:2006za} for showering and hadronization.
We adopted MLM matching scheme~\cite{Alwall:2007fs} for matrix element and parton shower merging.
The event samples are finally fed into fast detector simulator Delphes~3.4.0~\cite{deFavereau:2013fsa}
for detector effects. Here we have incorporated default ATLAS based detector card available within Delphes framework. 
We do not include  backgrounds from  the fake and non-prompt sources. 
Such backgrounds are not properly modeled in Monte Carlo simulations and requires data to estimate such contributions.

The LO $t\bar t+$jets and $Wt+$jets cross sections are normalized to NNLO+NNLL cross sections by  factors
1.84~\cite{twiki} and $1.35$~\cite{Kidonakis:2010ux} respectively. The DY+jets background cross section 
is adjusted to the NNLO QCD+NLO EW one by a factor 1.2, which is obtained utilizing FEWZ 3.1~\cite{Li:2012wna}.
The $t\bar t Z$,  $\bar tZ +$ jets, $t\bar t h$, $4t$ and  $t\bar{t} W^-$ ($t\bar{t} W^+$) cross sections at LO are
normalized to NLO ones by the  $K$-factors 1.56~\cite{Campbell:2013yla},
1.44~\cite{Alwall:2014hca}, 1.27~\cite{twikittbarh}, 2.04~\cite{Alwall:2014hca} and 1.35 (1.27)~\cite{Campbell:2012dh} respectively,
while $tWh$ and $tWZ$ are kept at LO. Further, the $W^-Z+$jets background is normalized to NNLO cross section by factor 
2.07~\cite{Grazzini:2016swo}. For simplicity, we assumed the 
QCD correction factors for the  $ tZj$ and $W^+Z+$jets to be the same as their 
respective charge conjugate processes. The signal cross sections are kept at LO.

\begin{table}[hbt!]
\centering
\begin{tabular}{|c |c| c| c| c | c| c| c |c |c|}
\hline
&&&&&&&&\\ 
      BP          & $t\bar t+$  & $DY+$      & $Wt+$    &  $t\bar t Z$   &  $t\bar t h$    & $t Z+$  & Others & \ Total  \          \\
                  &  jets       &jets       &jets       &                &                 & jets        &        &   Bkg.   \\      
&&&&&&&& (fb)\\                                
\hline
\hline
&&&&&&&&\\
       I            & 0.477      & 0.975       & 0.372     & 0.038        & 0.012             & 0.014    & 0.005  & 1.893   \\ 
       II           & 0.391      & 0.747       & 0.252     & 0.039        & 0.007             & 0.009    & 0.004  & 1.449            \\
       III          & 0.198      & 0.458       & 0.111     & 0.027        & 0.002             & 0.006    & 0.002  & 0.804 \\
\hline
\hline
\end{tabular}
\caption{The cross sections (in fb) for the different background contributions of 
the $bZH$ process after selection cuts at $\sqrt{s}=14$ TeV LHC.
The subdominant backgrounds $4t$, $t\bar t W$, $tWh$, $tWZ$ and $WZ$+jets are added together and denoted as 
``Others'' in the second last column, while the last column conforms the total background (Total Bkg.) yield. }
\label{bkgcompbA}
\end{table}
\begin{table}[hbt!]
\centering
\begin{tabular}{|c |c| c| c | c }
\hline 
                     BP & $|\rbb|$   &  \ Signal \           &  \ Significance ($\mathcal{Z}$)     \\ 
                        &          &       (fb)               &  300 fb$^{-1}$    \\                                      
\hline
\hline

                     I     &         & 0.548                    &  6.6    \\ 
                     II    & 0.1     & 0.286                    &  4.0    \\
                     III   &         & 0.119                    &  2.2     \\ 
\hline
\hline
\end{tabular}
\caption{The signal cross sections after selection cuts
of the $bZH$ process for the three benchmark points are presented in the third column for $|\rbb| = 0.1$. 
The corresponding significances are given in the fourth column.}
\label{signibA}
\end{table}

In order to distinguish the signal from the background processes, we have applied following
event selection criteria: Each event should contain two same flavor opposite sign leptons ($e$ and $\mu$), 
at least three jets with at least three of them are $b$-tagged. The minimum transverse momenta ($p_T$)
of the leading and subleading leptons are required to be $> 28$ GeV and $> 25$ GeV respectively, where as 
the $p_T$ of all three $b$-jets should be $> 20$ GeV. The absolute value of the pseudo-rapidity ($|\eta|$) of the leptons and all three $b$-jets
are needed to be $<2.5$. The jets are reconstructed by anti-$k_T$ algorithm with radius parameter $R = 0.6$.
The separations $\Delta R$ between any two $b$-jets, 
any two leptons and, any $b$-jet and lepton in an event are required be $> 0.4$.
In order to reduce the $t\bar t$+jets background we vetoed events with missing transverse energy ($E_T^{\mbox{miss}}$) $> 35$ GeV.
The invariant mass of the two same flavor opposite charge leptons ($m_{\ell\ell}$) is required to be within the $Z$ boson mass window 
i.e. $ 76 < m_{\ell\ell} < 100$ GeV. To reduce backgrounds further, 
we finally demanded the invariant mass of the two same flavor opposite charge leptons and two leading 
$b$-jets ($m_{\ell\ell bb}$) to remain within $|m_A- m_{\ell\ell bb}|< 100$ GeV. The normalized $m_{\ell\ell}$ and 
$m_{\ell\ell bb}$ distributions before application of any selection cuts are presented in Appendix.
We adopted the $\eta$ and $p_T$ dependent
$b$-tagging efficiency and, $c$- and light-jets misidentification efficiencies of Delphes. The background cross sections of the three benchmark points
after selection cuts are summarized in Table.~\ref{bkgcompbA}, while the signal cross sections along with their corresponding
significances with the integrated luminosity $\mathcal{L}= 300$ 
\fbi are given in Table~\ref{signibA} for $|\rbb| = 0.1$. We remark that in our exploratory
study we have not optimized the selection cuts such as $m_{\ell\ell}$ and $m_{\ell\ell bb}$,
and kept them unchanged for all three benchmark points, 
however impact of changing them will be discussed in the following.

The statistical significances in  Table~\ref{signibA} are determined by using
$\mathcal{Z} = \sqrt{2[ (S+B)\ln( 1+S/B )-S ]}$~\cite{Cowan:2010js},
where $S$ and $B$ are the number of signal and background events after selection 
cuts. The achievable significances
for BPI, BPII and BPIII are $\sim 6.6\sigma$, $\sim 4.0\sigma$ and $\sim 2.2\sigma$ with 300 \fbi integrated luminosity. 
We find that even the collected Run-2 data ($\sim150$ \fbi) would lead to $\sim 4.7\sigma$, $\sim 2.8\sigma$ 
significances for the BPI and BPII respectively, where as lower than $2\sigma$ for BPIII. 
As for the parameter space of EWBG, $|\irbb|$ should be $\gtrsim 0.058$~\cite{Modak:2018csw}, which 
leads to $\sim 12.1\sigma$, $\sim 4.5\sigma$ and $\sim2.5\sigma$ significances
for the BPI, BPII and BPIII with the full HL-LHC dataset (i.e. 3000 \fbi integrated luminosity). 
This implies that the $bZH$ process can fully probe
the parameter space required for $\rbb$ driven EWBG if $m_A \lesssim 300$ GeV, 
where as evidence ($3\sigma$) could be found for $300~\mbox{GeV}\lesssim m_A \lesssim 400$ GeV all three BPs. 
Here, we have kept the $Z$-pole invariant mass cut fixed to $76 ~\mbox{GeV} < m_{\ell\ell} < 100$ GeV.
The significances change negligibly up to $76 ~\mbox{GeV} < m_{\ell\ell} < 110$ GeV for all three BPs, however, reduce if broadened
further. On the other hand, we have kept $m_{\ell\ell bb}$ cut fixed to $|m_A- m_{\ell\ell bb}|< 100$ GeV for all three BPs. 
Significances do change with the change in $|m_A- m_{\ell\ell bb}|$ cut, however, mildly.  We have checked that, e.g., if the cut $|m_A- m_{\ell\ell bb}|$ is set to $< 80$ GeV
the significances enhance by $\sim 1\%$, $\sim 5\%$ and $\sim3\%$ for BPI, BPII and BPIII respectively, while reduce by 
$\sim2 \%$, $\sim3\%$ and $\sim2\%$ if set to $<120$ GeV. A narrower $m_{\ell\ell bb}$ cut than 80 GeV again reduces the significance for 
all three BPs.

Before closing, we remark that the scope for discovery of the $bZh$ process (i.e.
$pp\to b A + X \to b Z h +X$ with $Z \to \ell^+ \ell^-$, $h\to b \bar b$) is limited if $\cba$ is small. 
E.g., if $|\rbb|=0.1$, $\cba=0.05$, $m_A = 388$ and $m_H=236$~\footnote{Corresponding parameters for this scanned point are: 
$\eta_1=0.259,~\eta_2=0.838,~\eta_3=1.633,~\eta_4=-0.044,~\eta_5=-1.569,~\eta_6=-0.033,~m_{H^\pm}=323~\mbox{and}~\eta_7=0.130$.}, 
the significance lies below $\sim1\sigma$, even for the full HL-LHC dataset~\footnote{
In evaluating the significance we assumed the same cut based analysis as in $bZH$, except the application of an additional  
$|m_h - m_{bb}| < 25$ GeV cut.}. The significance improves substantially 
if $m_A < m_H +m_Z$ and/or $\rbb$ is large. A larger $\cba$ 
would also help, however in such cases the significance would be balanced by more severe
bounds from Higgs signal strength measurements.
For $\cba\sim 0.05$, $H \to Z Z$, $H\to W^+W^-$ would open up, 
but we do not find them to be very promising even for HL-LHC.

%%%%%%%%%%%%%%%%%%%%%%%%%%%%%%%%%%%%%
\subsection{The $bbZH$ process}
%%%%%%%%%%%%%%%%%%%%%%%%%%%%%%%%%%%%
As for the $bbZH$ process, the SM backgrounds are essentially the same 
as in the preceding subsection, however with one extra $b$-jet in the final state.
We have adopted similar procedure for the signal and background events generation and, followed
the event selection cuts as in the $bZH$ process except the additional 
$b$-jet is required to have $p_T > 20$ GeV and $|\eta| < 2.5$.  
The separation $\Delta R$ between any two $b$-jets, 
any two leptons and, any $b$-jet and lepton should be $> 0.4$. 
All other cuts are kept same as in $bZH$.
Finally, we applied the $m_{\ell\ell}$ and $m_{\ell\ell bb}$ selection cuts as before.
The background and signal cross sections after the selection cuts
are summarized in Table~\ref{bkgcompbbA} and Table~\ref{signibbA} respectively. 
We assumed the QCD correction factors for the different
backgrounds as in the $bZH$ process and kept the signal cross sections at LO. Therefore, we remark that, there
are slightly greater uncertainties involved in the background cross sections.

As can be seen from Table~\ref{signibbA}, the cross sections of the $bbZH$ process is suppressed due its to 
$2 \to 4$ body nature. Hence, the significances are provided only for 3000 \fbi integrated luminosity,
which can reach up to $\sim 4.2\sigma$, $\sim 2.8\sigma$ and $\sim 1\sigma$ for the BPI, BPII and BPIII respectively. 
Hence, a discovery is beyond the HL-LHC, that is unless $\rbb$ is large.
The significances can be higher if the upper limits of $|\rbb|$ for the corresponding BPs 
are saturated, which can rise up to $\sim 6.5 \sigma$, $\sim 13.8 \sigma$ and $\sim 4.5 \sigma$
for BPI, BPII and BPIII respectively with the full HL-LHC data.  
The impacts of other choices of $m_{\ell\ell}$ are similar as in $bZH$ process. 
However, if $|m_A- m_{\ell\ell bb}|$ cut is set to $<80$ GeV ($<120$ GeV), the significances enhance (reduce) 
by  $\sim 2\%$ ($\sim4\%$), $\sim 11\%$ ($\sim 5\%$) and $\sim 4\%$ ($\sim 2\%$) for  BPI, BPII and BPII respectively.
As in before, $pp\to b\bar b A + X \to b\bar b Z h + X$ is possible, however, the 
significances are even smaller than the $bZh$ process.

\begin{table}[hbt!]
\centering
\begin{tabular}{|c |c| c| c| c | c| c| c |c |c|}
\hline
&&&&&&&&\\ 
      BP          & $t\bar t+$  & $DY+$      & $Wt+$    &  $t\bar t Z$   &  $t\bar t h$    & $t Z+$  & Others & \ Total  \          \\
                  &  jets       &jets       &jets       &                &                 & jets        &        &   Bkg.  \\      
&&&&&&&& (fb)\\                                
\hline
\hline
&&&&&&&&\\
              I    & 0.013       & 0.065      & 0.031    & 0.002          & 0.003           & 0.0005      & 0.0002  & 0.115 \\
              II   & 0.016       & 0.064      & 0.018    & 0.001          & 0.002           & 0.0003      & 0.0002  & 0.102 \\
              III  & 0.011       & 0.039      & 0.018    & 0.001          & 0.0008          & 0.0002      & 0.0002  & 0.071 \\
 
\hline
\hline
\end{tabular}
\caption{The cross sections (in fb) of the different backgrounds for the $bbZH$ process after selection cuts at $\sqrt{s}=14$ TeV.}
\label{bkgcompbbA}
\end{table}
\begin{table}[hbt!]
\centering
\begin{tabular}{|c |c| c| c | c }
\hline
%&&&\\ 
                     BP & $|\rbb|$   &  \ Signal \         &  \ Significance ($\mathcal{Z}$)     \\ 
                        &          &       (fb)              &    3000 fb$^{-1}$    \\      
%&&& \\                                
\hline
\hline
                     I    &         & 0.027        &  4.2                                      \\ 
                     II   & 0.1     & 0.017        &  2.8                                       \\
                     III  &         & 0.005        &  1.0                               \\ 
\hline
\hline
\end{tabular}
\caption{Same Table as in Table~\ref{signibA} however for the $bbZH$ process. }
\label{signibbA}
\end{table}

%%%%%%%%%%%%%%%%%%%%%%%%%%%%%%%%%%%%%%%%%%%%%%%%%%%%%%%%%
\section{Discussion and summary}\label{summ}
%%%%%%%%%%%%%%%%%%%%%%%%%%%%%%%%%%%%%%%%%%%%%%%%%%%%%%%%%
Motivated by the recent observation of $hbb$ coupling, we have investigated the possibility of 
probing extra bottom Yukawa coupling $\rbb$ at the LHC. 
We first looked for the existing constraints on $\rbb$, mainly 
from the Higgs signal strength measurements, $\bsg$, $\dcp$ of $\ebsg$, electron EDM, as well as
several direct searches at the LHC. We found that $\mathcal{O}(0.1)$ $|\rbb|$ is allowed by the current data, however
$\cba$ and $\rtt$ should not be large. We remark that additional constraints  can come from the $\mathcal{A}_{\text{CP}}$ and
isospin violating asymmetry ($\Delta_{0+}$) of $B\to K^*\gamma$ measurement by Belle~\cite{Horiguchi:2017ntw}, and
could be comparable to the inclusive one, however, 
they both suffer from sizable uncertainties in their theoretical predictions~\cite{Hurth:2010tk}.

We have shown that \bZH  with $Z\to \ell^+ \ell^-$ and $H\to b \bar b$ offers 
excellent probe for $\rbb$. Discovery seems plausible with 300 \fbi integrated luminosity for $\mathcal{O}(0.1)$ $\rbb$ however 
$m_A$ needs to be $\lesssim300$ GeV. For $400~\mbox{GeV}\lesssim m_A\lesssim 500$ GeV one may need the HL-LHC data.
The process could be followed by $gg\to b \bar b A \to b \bar b Z H$, however, we find that a discovery is 
unlikely even with the full HL-LHC dataset if $|\rbb|\simeq 0.1$. 
We focused on the scenario where $m_H + m_Z < m_A$.
However, for $m_A + m_Z < m_H$ our study can be extended to $bg\to b H \to b A Z$ (and $gg\to b \bar b H\to b \bar b Z A$) process where
a complementary search strategy as in $bZH$ (and $bbZH$) can be adopted.
Note that, $\rbb$ also invokes $gg\to \bar t b H^+ \to \bar t b t \bar b$, which we leave out for future study.
We have not included QCD corrections for the signal and neglected systematic uncertainties in our analysis. 

As a first estimate we have not included uncertainties arising from factorization scale ($\mu_F$) and
renormalization scale ($\mu_R$) dependences in our LO cross section calculation. In general, LO $bg\to bA/bH$ cross section 
has sizable scale uncertainties ($\sim25-30\%$) in the mass range $m_{A/H}\sim 300-400$ GeV 
for bottom quark $p_T > 15-30$ GeV and $|\eta|<2.5$~\cite{Campbell:2002zm} (see also~\cite{Dicus:1998hs,Maltoni:2005wd}). 
The uncertainties are much higher for $gg\to b\bar b A/H$. E.g., the scale uncertainties could be as large as
$\sim40\%$ at LO~\cite{Harlander:2003ai} depending on the choices of $\mu_R$ and $\mu_F$. For both these 
processes the uncertainties are much smaller at NLO (or at NNLO)\cite{Campbell:2002zm,Harlander:2003ai}.
It was discussed in Ref.~\cite{Maltoni:2003pn} that the LO cross sections
estimated with LO PDF set CTEQ6L1~\cite{Pumplin:2002vw} have large factorization scale dependence.
We remark that our LO cross sections, which are determined 
utilizing LO PDF set NN23LO1, may have similar level of uncertainties.
It is found that at $\mu_F\approx m_{A}~(\mbox{or}~m_H)$  
the corrections are large negative ($\sim-70\%$)~\cite{Maltoni:2003pn}, while
at $\mu_F\approx m_A/4 ~(\mbox{or}~m_H/4)$ the corrections are small; indicating $\mu_F\approx m_A/4$ is the relevant
factorization scale. In particular, the cross section uncertainties from renormalization and 
factorization scale choices are found to be small for $\mu_R=m_A$ and varied between
$\mu_R=m_A/2$ to $\mu_R=2m_A$, along with $\mu_F = m_A/4$ and 
varied between $\mu_F=m_A/8$ to $\mu_F=m_A/2$~\cite{Maltoni:2003pn}.
Furthermore, we have not included PDF uncertainties, which could be substantial for the heavy quark initiated process,
as discussed e.g. in Ref.~\cite{Maltoni:2012pa}. 
A discussion on PDF choices and their uncertainties for LHC can be found out in 
Ref.~\cite{Butterworth:2015oua}.  These would induce some uncertainties in our results.
A detailed estimation of such uncertainties we leave out for future.

A discovery might indicate EWBG driven by $\rbb$. With the full HL-LHC dataset the $bZH$ process can probe 
the entire parameter space required for the EWBG if $m_A \lesssim 300$ GeV. 
Although, a discovery would be intriguing, however it would not be sufficient to establish it to the EWBG
without the information of the phase of $\rbb$. This would need further scrutiny 
and perhaps angular analysis of the $bZH$ (or $bbZH$) process
would be indicative. Information of the phase can also be extracted from the future measurement of $\dcp$ of $\ebsg$ at Belle-II, 
if $H^\pm$ is not too heavy.

In principle, $\rho_{bd}$, $\rho_{db}$, $\rho_{bs}$ and $\rho_{sb}$ all can 
replicate $bZH$ and $bbZH$ signatures at the LHC, however, their impacts are inconsequential
due to severe bounds from $B_d$ and $B_s$ mixings. If the charm quark gets
misidentified as $b$-jet, a sizable $\rho_{cc}$ can mimic $bZH$ signature in $pp$ collision via $cg\to c A \to c Z H$. 
However, such possibilities can be excluded with the simultaneous application of $c$- and $b$-tagging 
on the final state event topologies~\cite{Hou:2018npi}. 

While determining the discovery potential we set all $\rho_{ij}=0$ except for $\rbb$ for the sake of simplicity.
In general, non-zero $\rho_{ij}$s suppress $\mathcal{B}(A/H\to b \bar b)$ and hence the discovery 
potential of the $bZH$ and $bbZH$. It is plausible that $\rho_{ij}$ 
couplings might share the same flavor organization patterns of the SM,
i.e. trickling down off-diagonal elements as observed in the quark masses and mixings. If so,
$\rho_{ii}$ could be $\sim\lambda_i$ and, similarly $\rho_{\tau\tau}\sim \lambda_\tau$~\cite{Hou:2017hiw}.
In demonstrating the discovery potential of $bZH$ and $bbZH$ processes,
we have chosen $|\rbb| =0.1$ for illustration, 
however, $|\rho_{bb}|$ could be $\sim\lambda_b$. For $|\rho_{bb}|=\lambda_b$, one has $\sim3\sigma$ for BPI with the 
full HL-LHC data for the $bZH$ process, where as, significances lie below $1\sigma$ for the BPII and BPIII.
We remark, $\rho_{\tau\tau}\sim \lambda_\tau$
impacts negligibly on the discovery potential. 
Throughout we assumed the flavor changing neutral Higgs coupling $\rho_{tc}$ to be small, 
however, a $\mathcal{O}(1)$ value is still allowed by the current 
data~\cite{Kohda:2017fkn,Hou:2018zmg} (see also Ref.~\cite{Altmannshofer:2019ogm}), and
could potentially reduce the significances of both the processes. 
  
We assumed small $\rtt$ in order to avoid strong constraints arising from the $gg\to A/H$ searches.
Notwithstanding, $\mathcal{O}(1)$ $\rtt$ with complex phase provides another robust 
mechanism for EWBG~\cite{Fuyuto:2017ewj} (see also Ref.~\cite{deVries:2017ncy}), 
which can be probed by the conventional search programs such as $gg\to A/H\to t \bar t$ 
or $gg\to A/H t \bar t\to t \bar t t \bar t$~\cite{Craig:2016ygr}.
The former process suffers from large interference~\cite{Carena:2016npr} with the overwhelming $gg\to t\bar t$ background, 
however recent ATLAS~\cite{Aaboud:2017hnm} and CMS~\cite{CMS:2019lei} studies found some sensitivity. We remark that 
$\rtt$ could be $\sim\lambda_t$. Utilizing results from Refs.~\cite{CMS:2019nig,Aaboud:2018cwk,CMS:1900zym},
Ref.~\cite{Hou:2019gpn} found $|\rtt|\gtrsim 0.8-0.9$ are excluded (at 95\% CL) if $m_A/m_H \approx 400-500$ GeV, 
or $m_{H^{\pm}}\sim 300-400$ GeV. 
These are roughly the ballpark values of $m_A$ and $m_{H^{\pm}}$ of the three BPs, as can be seen from Table~\ref{bench}.
Non-zero $\rtt$, alleviates the discovery potential of both $bZH$ and $bbZH$ processes via $\mathcal{B}(A/H\to t\bar t)$,
for $m_A/m_H > 2 m_t$. For example, taking $|\rtt|=0.8$ as yardstick and by a simple rescaling of the numbers
in Table~\ref{signibA} and Table~\ref{signibbA}, we find that the significances for BPII and BPIII drop by $\sim 60\%$ and $\sim70\%$ respectively. 
Therefore, discovery of $bZH$ is possible only for BPII (and BPI since $m_A$ lies below $2 m_t$ threshold),
but would require full HL-LHC data. In such cases, discovery is not possible for BPII and BPIII for $bbZH$ process, even with full HL-LHC data.
For sizable $\rbb$ and $\rtt$, $bg \to b A/H \to b t \bar t$ as well as $gg\to b \bar b A/H \to b \bar b t \bar t $~\cite{ttbb} 
both are possible, and would provide complementary information.

In summary, we have explored the possibility of discovery and identification of additional bottom Yukawa coupling that might exist in nature
via \bZH and \bbZH  processes at $\sqrt{s}=14$ TeV LHC. We found that the former process
could be discovered with 300 \fbi integrated luminosity if $m_A\sim 300$ GeV, which could be extend up to
$\sim 500$ GeV but the full HL-LHC dataset would be required. The latter process could also be discovered at the HL-LHC, however $\rbb$
needs to  be large. A discovery would not only confirm physics beyond the Standard Model, but may also 
indicate the EWBG driven by $\rbb$.

\begin{figure*}[htbp!]
\center
\includegraphics[width=.48 \textwidth]{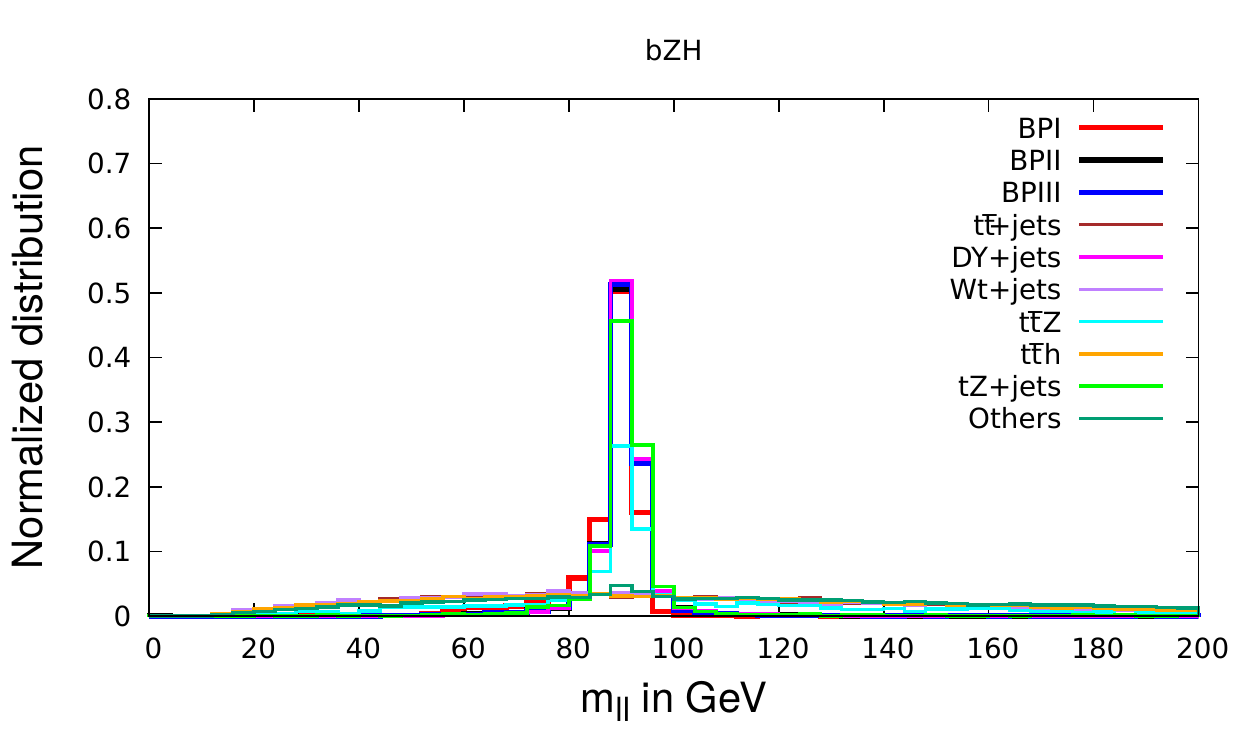}
\includegraphics[width=.48 \textwidth]{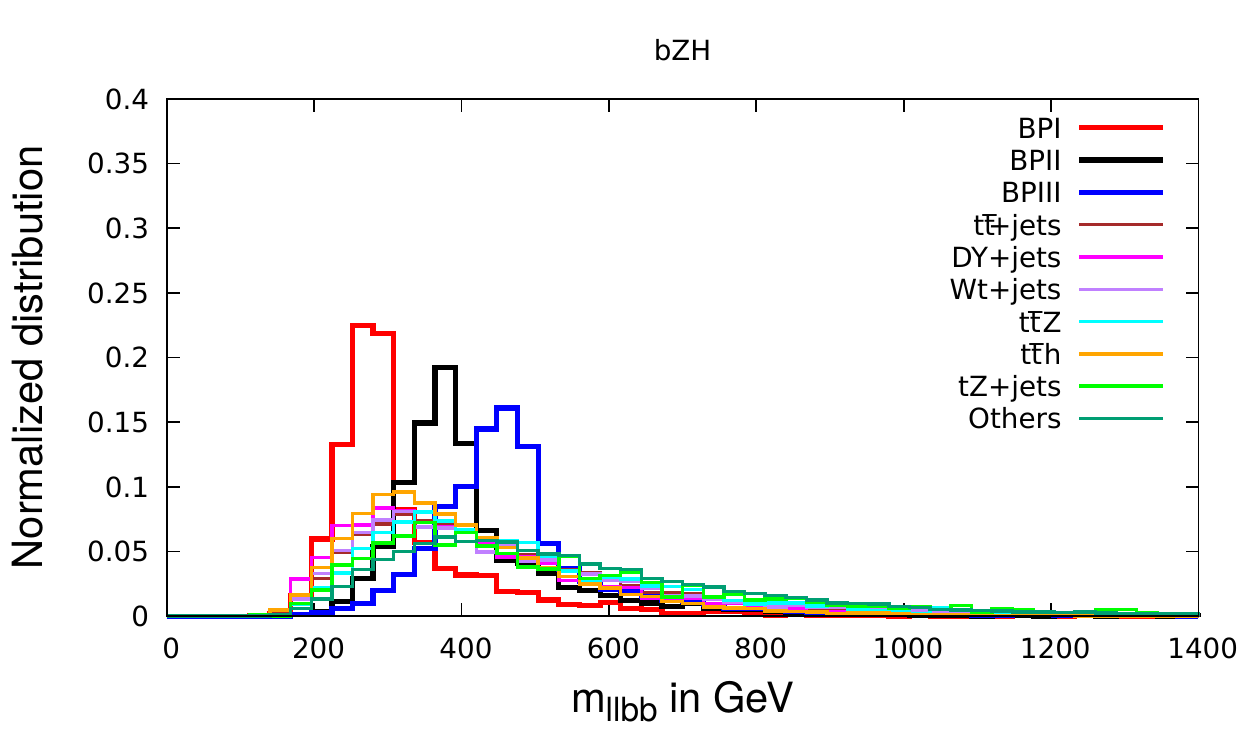}
\includegraphics[width=.48 \textwidth]{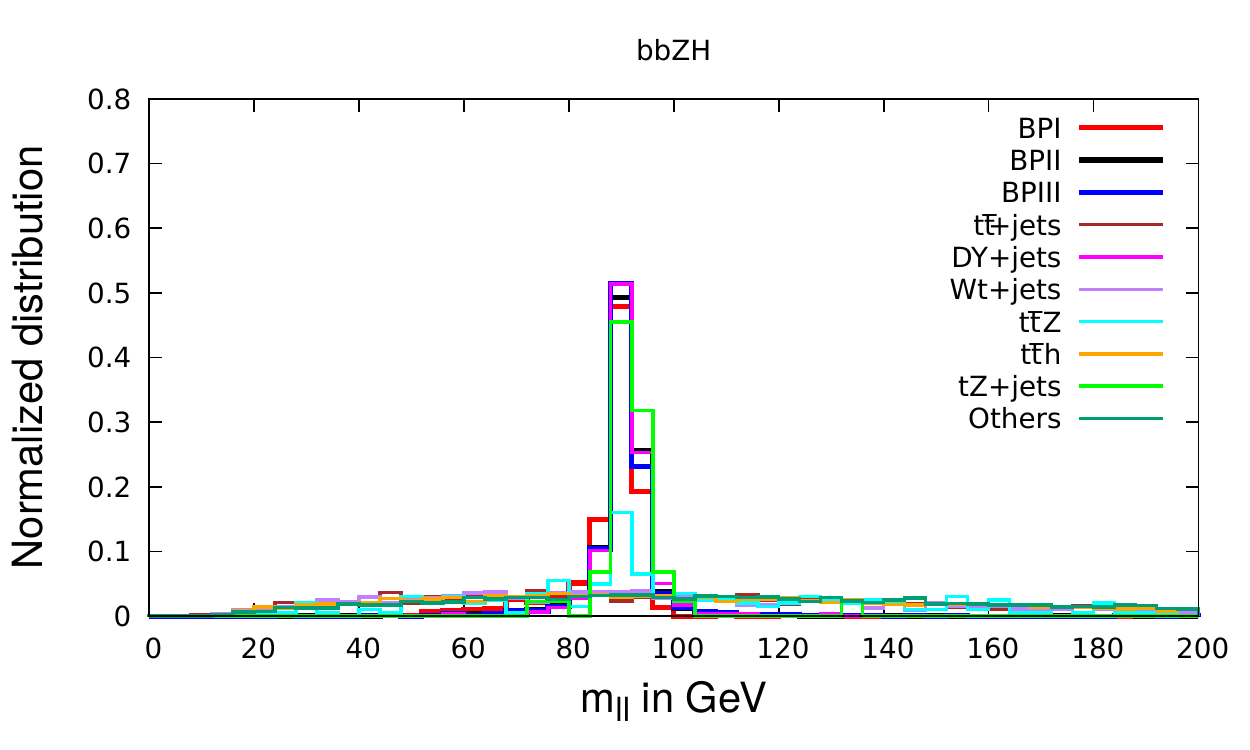}
\includegraphics[width=.48 \textwidth]{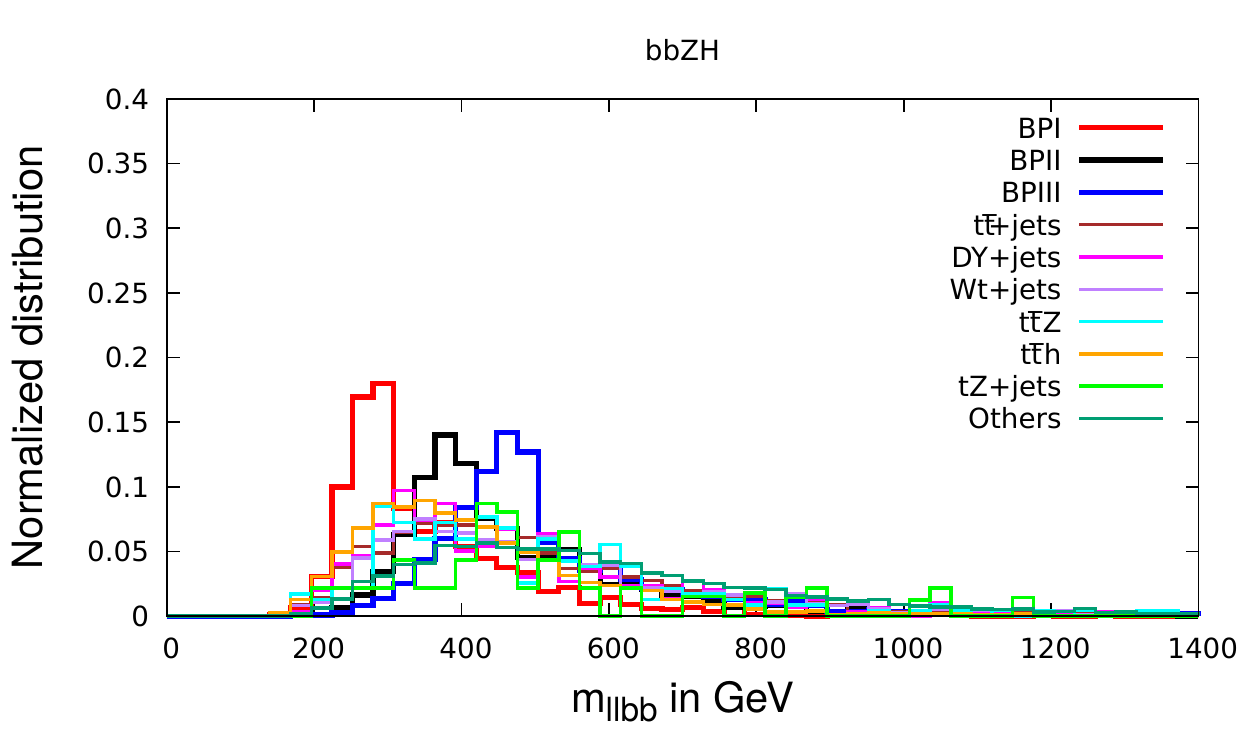}
\caption{The normalized $m_{\ell\ell}$ (left) and $m_{\ell\ell bb}$ (right) distributions of the three BPs together with
different backgrounds for $bZH$ (upper panel) and $bbZH$ (lower panel) processes.} 
\label{invmass}
\end{figure*}
\appendix

\vskip0.2cm
\begin{acknowledgments}
We thank  W.-S Hou, M. Kohda and E. Senaha for many fruitful discussions. We also
thank U.K. Dey for comments. This research is supported by grant MOST-107-2811-M-002-3069.
\end{acknowledgments}

\section{Invariant Mass Distributions}
\label{dist}

The normalized invariant mass distributions $m_{\ell\ell}$ and $m_{\ell\ell bb}$ for 
the signal and backgrounds of the  $bZH$ and $bbZH$ processes
are presented in Fig.~\ref{invmass}. These figures are generated 
without any selection cuts. The events for the signal and background processes are 
generated with the default cuts of MadGraph5\_aMC@NLO with minimal 
modifications, followed by showering and hadronization via PYTHIA~6.4 and, with ATLAS 
based Delphes card. 

%--------------------------------------------------------------

\end{document}